\documentclass[twocolumn,showpacs,preprintnumbers,amsmath,amssymb,prb,superscriptaddress]{revtex4}

\pdfoutput=1

\usepackage{graphicx}% Include figure files
\usepackage{dcolumn}% Align table columns on decimal point
\usepackage{bm}% bold math
\usepackage{amsmath}
\usepackage{amssymb}
\usepackage{color}

\newcommand{\CQCTsyd}{Centre of Excellence for Quantum Computer Technology, School of Electrical Engineering \& Telecommunications,
\\ University of New South Wales, Sydney NSW 2052, Australia.}
\newcommand{\CQCTmel}{Centre of Excellence for Quantum Computer Technology, School of Physics,
University of Melbourne, Melbourne VIC 3010, Australia.}
\newcommand{\TKKa}{Department of Applied Physics/COMP, Helsinki University of
  Technology, P.O.~Box 5100, FI-02015 TKK, Finland.}
\newcommand{\TKKb}{Low Temperature Laboratory, Helsinki University of Technology,
  P.O.~Box 3500, FI-02015 TKK, Finland.}

\begin{document}
\title{Transport Spectroscopy of Single Phosphorus Donors in a Silicon Nanoscale Transistor}

\author{Kuan~Yen~Tan}
\email{kuan-yen@student.unsw.edu.au} \affiliation{\CQCTsyd}

\author{Kok~Wai~Chan}
\affiliation{\CQCTsyd}

\author{Mikko~M\"{o}tt\"{o}nen}
\affiliation{\CQCTsyd} \affiliation{\TKKa} \affiliation{\TKKb}

\author{Andrea~Morello}
\affiliation{\CQCTsyd}

\author{Changyi~Yang}
\affiliation{\CQCTmel}

\author{Jessica~van~Donkelaar}
\affiliation{\CQCTmel}

\author{Andrew~Alves}
\affiliation{\CQCTmel}

\author{Juha-Matti~Pirkkalainen}
\affiliation{\CQCTsyd} \affiliation{\TKKa}

\author{David~N.~Jamieson}
\affiliation{\CQCTmel}

\author{Robert~G.~Clark}
\affiliation{\CQCTsyd}

\author{Andrew~S.~Dzurak}
\affiliation{\CQCTsyd}

\date{\today}
\newpage

\begin{abstract}

We have developed nano-scale double-gated field-effect-transistors for the study of electron states and transport properties of single deliberately-implanted phosphorus donors. The devices provide a high-level of control of key parameters required for potential applications in nanoelectronics. For the donors, we resolve transitions corresponding to two charge states successively occupied by spin down and spin up electrons. The charging energies and the Land\'{e} g-factors are consistent with expectations for donors in gated nanostructures.

\end{abstract}

\pacs{73.21.-b,61.72.Vv}% PACS, the Physics and Astronomy

\maketitle

The ability to manipulate and measure electrons bound to
phosphorus (P) donors is a key ingredient for the realization of
quantum information processing schemes using single
dopants\cite{KaneNature98,hollenberg06PRB}. Recent development of
metal-oxide-semiconductor(MOS)-compatible nanostructures in
silicon\cite{LeeAPL06,FujiwaraAPL06,AngusNano07,LiuAPL08} together with single-ion detection
capabilities\cite{JamiesonAPL05} hold promise for the realization
of dopant-based spin qubits in silicon\cite{MorelloPRB09}. Such
qubits are attractive due to the long coherence times of donor
electron spins in bulk\cite{TyryshkinPRB03}, but further research
is necessary to understand how the donor spin coherence is
influenced by the proximity of a Si--SiO$_2$
interface\cite{schenkel06APL,desousa07PRB}. An important step towards
this goal is to create gated nanostructures where the energy
spectrum of individual donors can be observed and, possibly,
manipulated by changing the local environment or the position of
the donors with respect to the Si--SiO$_2$ interface. Pioneering
studies of dopant spectroscopy have been carried out on a variety
of
nanostructures\cite{SellierPRL06,LansbergenNatPhys08,CalvetPRB08,CalvetPRL07,CalvetPRB07},
but the presence and location of the donors were uncontrolled. Here, we describe a double-gated nanoscale
field-effect-transistor (nanoFET) where a chosen mean number of phosphorus donors
are deliberately implanted into the conduction channel, and
use this device to study the charge and spin states of these
individual P atoms. This structure allows convenient control of four key device
parameters: (i) gate-tunable electron density in the source and drain
reservoirs\cite{AngusNano07}; (ii) number of implanted donors\cite{JamiesonAPL05}; (iii) depth of implanted
donors; and (iv) tunnel coupling between
donors and reservoirs. In previous studies only the tunnel coupling was controlled by choosing the FET channel length.

\begin{figure}[th!] \center
%\texttt{.pdf}
\includegraphics[width=8.6cm]{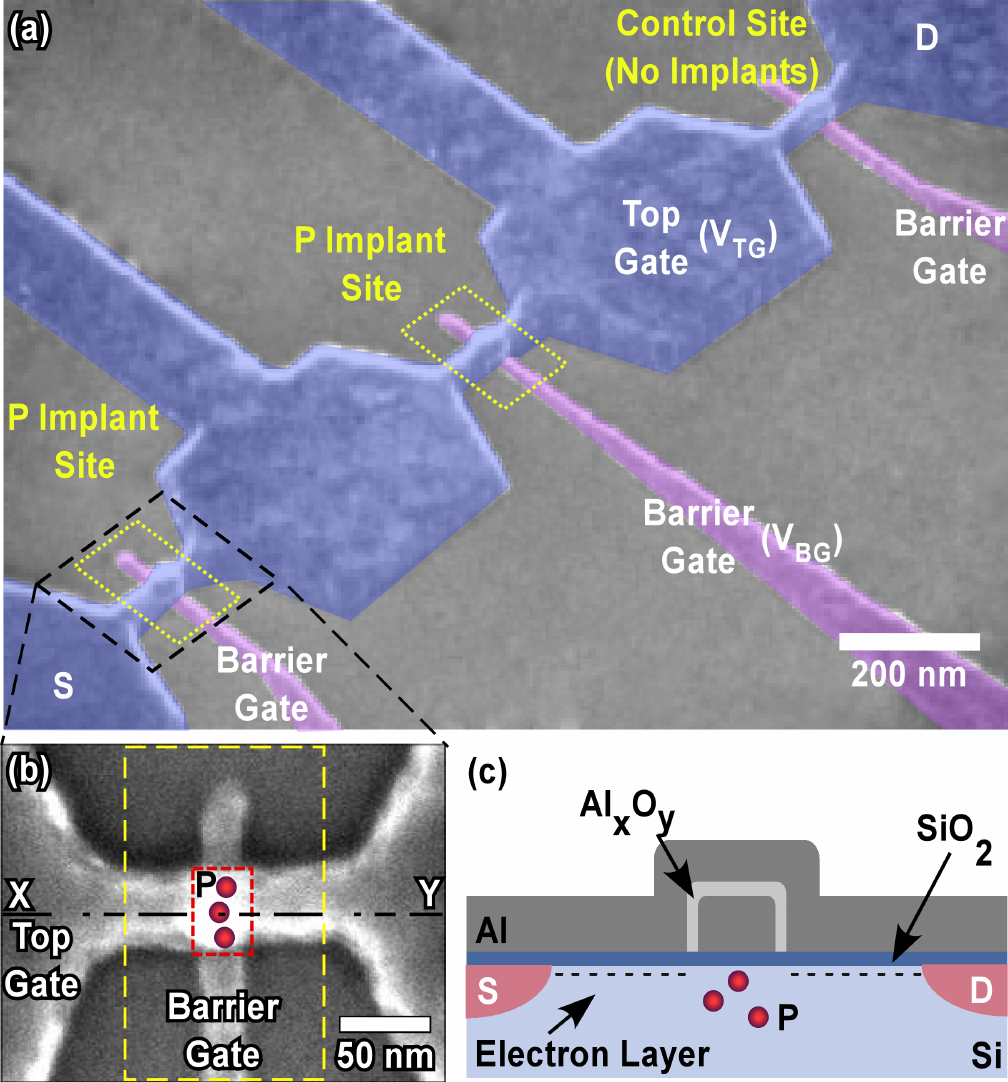}
\caption{ \label{tribarrier} (Color online) (a) Coloured SEM image
of a tri-barrier nanoFET. Positive voltage on the top gate
induces an electron layer which
extends $\sim$25~$\mu$m to source (S) and drain (D) ohmic contacts. Barrier gates, which
are partially underneath the common top gate, allow electrical
control of the donor energy levels, for example, the chemical
potential can be brought into resonance with Fermi levels of
adjacent reservoirs. (b) Zoomed-in SEM image of a single nanoFET device. Yellow dashed line indicates the P implant window and red
dotted line indicates the active region where the mean number of
donors is three.
(c) (Not to scale) Schematic cross section of a nanoFET along line XY in panel (b). The red source and drain regions are formed by n$^{+}$ diffused contacts.}% end caption
\end{figure}
The nanoFET studied in this paper [see Figure~\ref{tribarrier}(c)] consists
of two independent aluminum gates: a top gate that is biased
positively to induce an electron layer that constitutes the source
and drain reservoirs; and a barrier gate that depletes the electron
layer in the active region of the donors, creating tunnel barriers
on each side of the donor potential well. This structure enables the
electrical manipulation of the charge states of the individual
phosphorus donors under the barrier gate. In particular, the barrier
gate voltage ($V_{\textrm{BG}}$) can be used to tune the donor
electrochemical potentials into resonance with those of the source
and drain reservoirs, inducing electron transport. The electron
density in the source and drain reservoirs can be tuned
\textit{in-situ} by the top gate voltage ($V_{\textrm{TG}}$), while
the lithographic width of the barrier gate influences the width, and
therefore the transparency, of the tunnel barriers between donor and
reservoirs. The excitation spectrum of the donor and its magnetic
field dependence extracted from the transport measurements provide
important information on electronic properties of donors in close
proximity to gate electrodes and induced electron layers.

The devices were fabricated on a high-resistivity ($\sim$10
k$\Omega$cm) near-intrinsic natural-isotope silicon (100)
wafer\cite{Topsil}, with low residual P background doping
($\sim$10$^{12}$ cm$^{-3}$). Ohmic contacts for the n$^{+}$ source
and drain are phosphorus-diffused regions fabricated using standard
UV-lithography and thermal diffusion at temperatures $\sim$
950~$^{\circ}$C. The 5~nm gate oxide is high-quality SiO$_{2}$ grown
by dry thermal oxidation. A 5~s rapid thermal anneal (RTA) at
1000~$^{\circ}$C N$_{2}$ ambient was carried out to lower the
interface trap density to the 2$\times$10$^{10}$~cm$^{-2}$eV$^{-1}$
range, as measured on similarly processed
chips\cite{McCallumMRSS08}. For the spatially-selective P
implantation, we used a 150~nm polymethylmethacrylate (PMMA) layer
deposited above the SiO$_{2}$ as a mask, with 100~nm $\times$ 200~nm
apertures defined using electron beam lithography (EBL). In the
implantation process, ion acceleration energies of 14~keV and 10~keV
were utilized and the flux of the P ions was controlled to give an
average of $N_{d}\sim3$ individual P donors in the 50~nm $\times$ 30~nm
active area of the device, that is, the region below the overlap of
the top and barrier gates [red box in Figure~\ref{tribarrier}(b)].
Subsequently, another RTA at 1000 $^{\circ}$C for 5 s was applied to
activate the donors and repair any damage caused by the implantation
process.

The barrier gates used to control the energy levels of the P donors
and to locally deplete the electron layer were fabricated using EBL,
thermal evaporation of aluminum, and liftoff. An insulating
$\sim$5~nm Al$_{x}$O$_{y}$ layer was then created using plasma
oxidation, where the barrier gates were exposed to a low-pressure
oxygen plasma ($\sim$150~mbar) for 3~min at 150~$^{\circ}$C. This
Al$_{x}$O$_{y}$ provides electrical insulation between the barrier
gates and top gate for voltage differences greater than $\sim
4$~V\cite{HeijPhD01,AngusNano07,LimAPL09}. A second layer of
aluminum forming the top gate was again defined using EBL, thermal
evaporation, and liftoff. Figure~\ref{tribarrier} shows a device similar to
the ones studied here and its schematic cross-section. Note that
there are three independently contacted barrier gates in each device
corresponding to three independent devices, for which individual
measurements can be done, thus increasing the device yield. The top
gate is common to all the barrier regions and the widening of the
top gate between barriers ensures that there are no accidental
quantum dots created. A useful feature of our structures is that,
when preparing the apertures for ion implantation, we can choose to
mask the areas around some of the barrier gates, thus obtaining
control devices where we know with certainty that no P donors are
present [see Figure~\ref{tribarrier}(a)].

We measured 27 P-implanted and 11 control devices altogether and
present here results from three different devices---one non-implanted control device (Sample A),
and two implanted (Samples B and C)---that represent
well the qualitative features of the whole batch. Samples B and C
differ in the energy used for the P ion implantation (14~keV in B,
10~keV in C) and in the width of the barrier gate (30~nm in B,
50~nm in C). All the experiments presented here were performed in dilution
refrigerators at base temperatures $T\lesssim$ 100~mK.

\begin{figure}[tbh!] \center
\includegraphics[width=8.6cm]{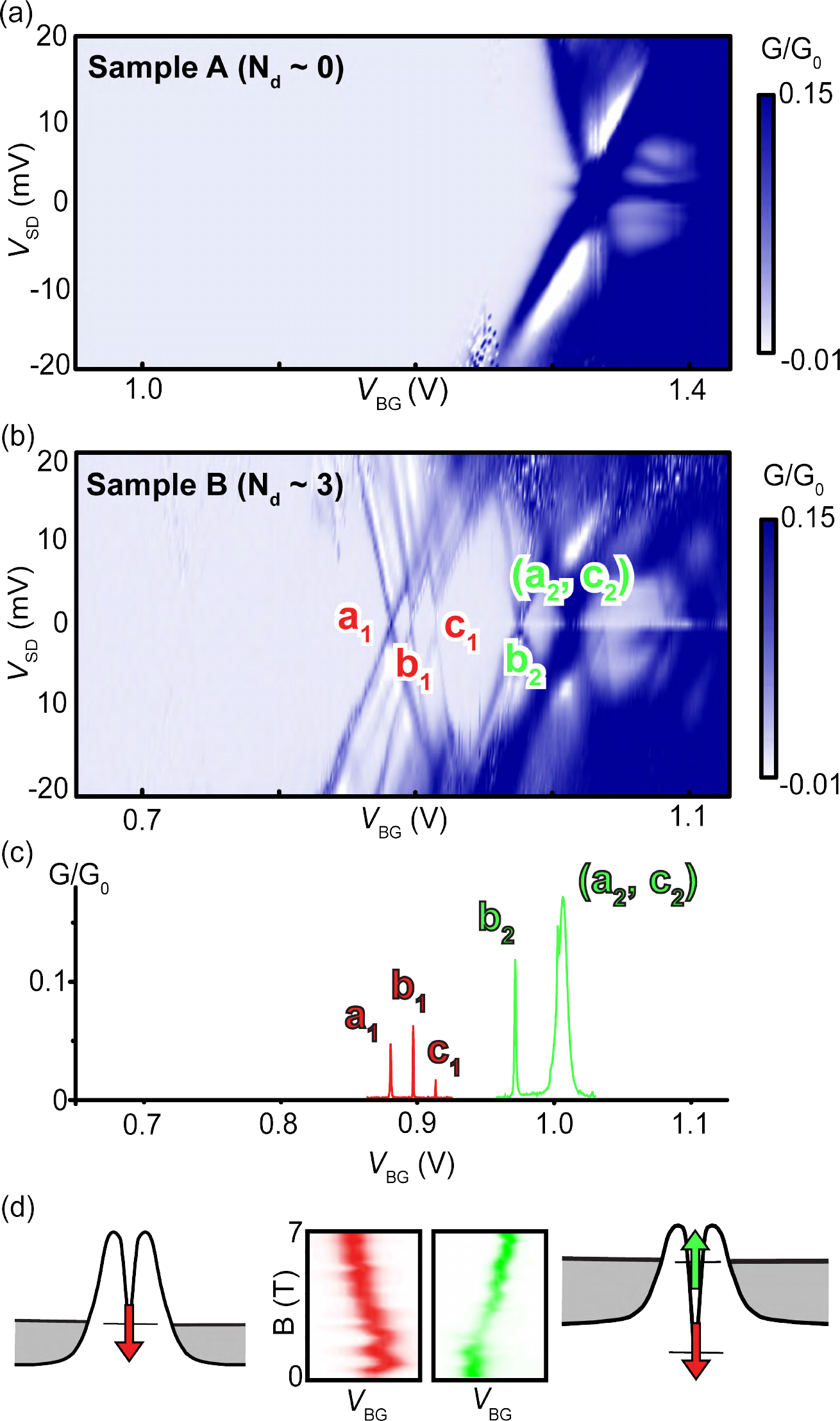}
\caption{\label{stabdiag}(Color online) (a) Stability diagram
showing differential conductance as a function of barrier gate and
dc source-drain (SD) bias of a control device (Sample A) with no
implanted donors.
(b) Stability diagram of Sample B with six labeled peaks.
(c) A cut of the stability diagram in panel (b) at $V_{\textrm{SD}}$ = 0~V, highlighting the resonant tunneling peaks due to three donors.
(d) Direction of peak shifts under global magnetic field up to 7 T. \textcolor{black}{Peak $a_{1}$ (red)} shifted with the field to the left while peak $b_{2}$ (green) shifted to the right. Schematics indicate that $a_{1}$ \textcolor{black}{is} due to tunneling of spin-down electrons and $b_{2}$ is a two-electron singlet state. All data in this figure were measured at $T\lesssim$~100~mK \textcolor{black}{, and \textit{$G_{0}=e^{2}/h$}.}}% end caption
\end{figure}

Figure~\ref{stabdiag}(b) shows the source-drain differential conductance of
Sample B as a function of the dc source-drain voltage and the
barrier gate voltage, with the top gate voltage fixed at
$V_{\textrm{TG}}$ = 3.5 V. We observe the channel turn-on as a
function of $V_{\textrm{BG}}$ when the tunnel barrier between source
and drain becomes transparent. In addition, sharp conductance peaks
are clearly visible at lower barrier voltages below the threshold.
These features are signatures of resonant tunneling through discrete
energy levels below the conduction band. Two sets of three different
peaks at 0.86~V$<V_{\textrm{BG}}<$0.93~V and 0.96~V$<V_{\textrm{BG}}<$1.03 V were observed [Figure~\ref{stabdiag}(b)]. These findings are consistent with the expected number of donors
($N_{d}\sim3$) in the active region. However, in other devices we often
found only one or two sets of sub-threshold conductance peaks. This
is consistent with the Poisson statistics related to the
implantation process\cite{Poisson}. In addition, some of
the donors can be off-center in the active region, and couple too
strongly to the source (drain) reservoir and too weakly to the drain
(source) reservoir. The total conductance is then dominated by the
slowest tunnel rate, making certain donors immeasurable in transport
even if they are physically present in the channel.

We identify peaks $a_{1}$, $b_{1}$, and $c_{1}$ as corresponding to
the transition from the positively charged state $D^{+}$ to the
neutral state $D^{0}$ of each donor, whereas peaks $a_{2}$, $b_{2}$,
and $c_{2}$ correspond to a transition to the charged state $D^{-}$.
The peaks associated with the same donor have almost the same lever
arm factors $\alpha = C_{\textrm{BG}}/C_{\Sigma}$, where
$C_{\Sigma}=C_{\textrm{BG}}+C_{\textrm{S}}+C_{\textrm{D}}+C_{\textrm{TG}}+C_{\textrm{0}}$
and $C_{\textrm{BG}}$, $C_{\textrm{TG}}$, $C_{\textrm{S}}$, and
$C_{\textrm{D}}$ are the coupling capacitances between the
corresponding donor and the barrier gate, top gate, source, and
drain, respectively. The self-capacitance of the donor is denoted by
$C_{\textrm{0}}$. This $\alpha$-factor can be determined from the
positive and negative slopes of each conductance peak in the
differential conductance plot [Figure~\ref{stabdiag}(b)], where the
positive slope is given by
$C_{\textrm{BG}}/(C_{\Sigma}-C_{\textrm{S}})$ and the negative slope
equals $C_{\textrm{BG}}/C_{\textrm{S}}$ (with drain
grounded)\cite{SellierPRL06}.

Figure 2(b) shows that the slopes of peaks $b_{1}$ and $b_{2}$ are
similar to each other, with $\alpha$ = 0.45, and hence they are
attributed to the same donor. Multiplying $\alpha$ by the difference
in barrier gate voltage $\Delta V_{\textrm{BG}}$ = 62~mV between
$b_{1}$ and $b_{2}$ yields the $D^{0}$ to $D^{-}$ charging energy
for donor $b$, $E_{c}$ = 28$\pm$3~meV\cite{FN1}. Peaks $a_{2}$ and
$c_{2}$ are just barely resolvable in Figure~\ref{stabdiag}(c), so we
cannot make a specific match with either of the $D^{0}$ peaks
$a_{1}$ and $c_{1}$. However, we can use the slope $\alpha$ = 0.26
of peak $a_{1}$, together with the gate voltage difference $\Delta
V_{\textrm{BG}}$ = 131~mV between peaks $a_{1}$ and
($a_{2}$,$c_{2}$), to calculate the charging energy for donor $a$ to
be $E_{c}$ = 34$\pm$6~meV. In both cases $E_{c}$ is lower than the
44 meV expected for phosphorus in
bulk\cite{RamdasRPP81,TaniguchiSSC76} due to the fact that there is
a substantial capacitive coupling of the donor to the surrounding
electrodes\cite{SellierPRL06,LansbergenNatPhys08}. Across the entire
series of implanted devices we have observed $E_{c}$ ranging from
21~meV to 39~meV, with an average value of 28~meV. A simple model of
the gate electrodes and the donor at the mean implantation depth in
silicon\cite{Fastcap,SRIM} provides a rough estimate of the
capacitances required to calculate the charging energy $E_{c} =
e^{2}/C_{\Sigma}$. With a donor depth $d$ = 5--20~nm below the
Si--SiO$_{2}$ interface, the model yields $E_{c}$ = 39--40~meV. The
discrepancy with the measured values can be, for example, due to the
limitations of a simple capacitance model for this ultra-small
few-electron system.

Magnetospectroscopic studies of Sample B are shown in
Figure~\ref{stabdiag}(d), obtained by applying a magnetic field up to 7~T in the plane of the electron layer. As the magnetic field was
increased, peaks \textcolor{black}{$a_{1}$ and $c_{1}$} shifted to lower
values of $V_\textrm{BG}$ whereas the peak $b_{2}$ shifted to higher
values\textcolor{black}{\cite{FN4}}. The direction of these shifts provides information about the
spin-polarity, relative to the external field, of the electrons that
contribute to the tunneling current: a shift to lower  (higher)
$V_\textrm{BG}$ corresponds to tunneling of spin-down (-up)
electrons. The magnetic-field-dependent shifts are consistent with
the assignment of peaks \textcolor{black}{$a_{1}$ and $c_{1}$ to $D^{0}$ and
$b_{2}$} to $D^{-}$ states, as inferred from the
bias spectroscopy in Figure~\ref{stabdiag}(b). This is due to the fact that
electrons in $D^{-}$, being a two-electron ground state, form a
singlet with opposing electron spins\cite{LossPRL01}. A quantitative analysis of the magnetic field
dependence is given for Sample C below.

As a control, another nanoFET structure (Sample A) was fabricated on
the same substrate but
with no implanted dopants in the active region. The stability
diagram is shown in Figure~\ref{stabdiag}(a), with $V_{\textrm{TG}}$ = 2.0~V. \textcolor{black}{It is significantly different from the data obtained for devices with donors in the active region. In particular, there are no sharp resonant tunneling features with large charging energies well below threshold. A weak minimum is observed in the threshold region (1.32~V$<V_{\textrm{BG}}<$1.40~V) which is most likely due to the formation of an unintentional dot, due to weak local disorder in the barrier region.
This is consistent with the small charging energy ($<$10~meV) and the large width of the peaks, which implies transparent tunnel barriers. Features reminiscent of Kondo-enhanced conductance\cite{GordonNat98,HauptmannNatPhys08,Rogge09} in this region, plus the absence of other peaks at lower $V_{\textrm{BG}}$, suggest that this could be the first electron in such a dot. The control data reinforce the interpretation that the peaks observed in Sample B are due to resonant tunneling through the deliberately implanted P donors.} We note that some
control devices processed without a final forming gas anneal (FGA)
showed sub-threshold conductance features which could be attributed
to localized states associated with interface traps, for example.
These features were clearly distinguishable from donors due to their
much lower charging energies, typically only up to 13~meV. We found
that a FGA can reduce significantly the charge trap density as well
as the charge noise\cite{AngusNano07}.

\begin{figure}[tbh!] \center
\includegraphics[width=8.6cm,height=15cm]{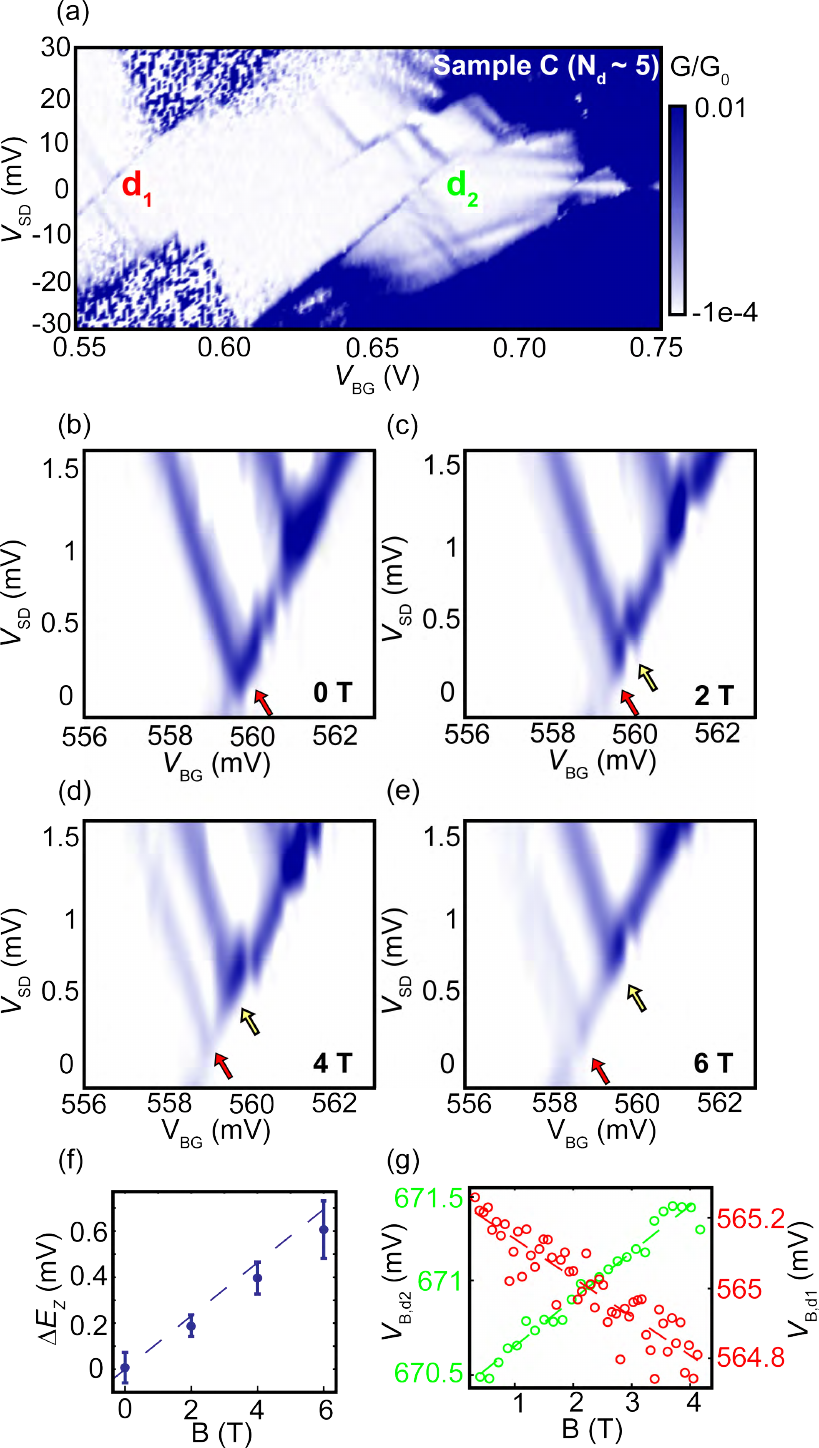}
\caption{ \label{fig3} (Color online) (a) Bias spectroscopy
showing differential conductance as a function of barrier gate and
dc SD-bias of Sample C. (b)--(e) Detailed bias spectroscopies of
peak $d_{1}$ at $B$ = 0, 2, 4, and 6 T. Red (yellow) arrows
indicate spin-down (-up) states entering the bias window. (f)
Energy splitting ($\Delta$$E_{Z}$) of the spin-up and spin-down states deduced from
panels (b)--(e). The dashed line indicates $2\mu_B B/e$ for
reference, where $\mu_B$ is the Bohr magneton. (g) Positions of
the peaks $d_{1}$ (red circles) and $d_{2}$ (green circles) in the
barrier gate voltage as a function of global magnetic field. The
dashed lines are linear fits to the data points.}% end caption
\end{figure}

Sample C was fabricated using the same processes as Sample B, with
the exception that the phosphorus donors were implanted with 10~keV
ion-beam acceleration energy, and the barrier gate was 50~nm wide,
yielding weaker tunnel coupling between each donor and the
reservoirs, and accordingly sharper conductance features. With 10~
keV implantation energy, the donors are expected to be $\sim$12~nm
from the Si--SiO$_{2}$ interface\cite{SRIM}, thus closer to the gate
compared with the donors implanted with 14~keV beam energy in Sample
B. The bias spectroscopy of Sample C is presented in Figure~\ref{fig3}(a).
Here, only one pair of $D^{0}$ and $D^{-}$ states, denoted by
$d_{1}$ and $d_{2}$, is strongly visible although the mean number of
donors under the active region (here $50\times 50$~nm$^2$) is
$N_{d}\sim5$. This is not surprising because the wider tunnel barriers,
as compared with Sample B, make the requirement of the donors being
well-centered even more stringent\cite{FN3}.

In the Coulomb blockaded region between $d_{1}$ and $d_{2}$,
additional faint conductance features were observed at finite
source-drain bias. These features are attributed to nearby weakly
coupled donors or other charge traps\cite{Pierre08}. We also
observed random telegraph signals (RTS) in the conductance in
well-defined regions at $\mid V_\textrm{SD} \mid$ $>$ 10 mV. This
phenomenon most likely arises because the electrochemical
potential (and thus the conductance) of the visible donor depends on
the charge state of the surrounding donors or traps, due to mutual
charging energy effects. Random switching of weakly coupled charge
centers can then give rise to the observed RTS. Thus the donor can
be possibly used as a charge sensor for the nearby charge
centers\cite{XiaoNat04}. Our preliminary theoretical analysis
shows that the distance between the visible donor and the nearby
donor or charge trap is more than 25~nm\cite{FN2}.

The sharpness of the conductance peaks in the wide-barrier Sample C
allowed us to study in more detail the spin states of the donor.
Figures 3(b--e) show a set of bias spectroscopy data on the peak
$d_{1}$ of Sample C, with applied magnetic fields $B = 0, 2, 4 ,
6$~T. For $B > 2$~T we clearly resolve an excited state at finite
bias, which we attribute to the excited Zeeman state (spin-up) of
the first electron on the donor. The peaks in the differential
conductance due to spin-down and -up electrons are indicated by red
and yellow arrows, respectively. The separation of these peaks on
the $V_{\rm BG}$~axis multiplied by $\alpha$ should then equal the
Zeeman splitting of the $D^0$ spin states. To verify this, we fitted
Gaussian functions to the differential conductance peaks in Figures
3(b--e) at various fixed SD bias points to obtain an accurate
measure of their position. The resulting energy splitting $\Delta
E_{Z}$ is plotted as a function of magnetic field in Figure~\ref{fig3}(f),
together with error bars obtained from the fitting procedure and the
error of $\alpha$ (including $\Delta E_{Z}$ at $B = 0$). As shown in
Figure~\ref{fig3}(f), $\Delta E_{Z}$ is in good agreement with the Zeeman
splitting for a Land\'{e} $g$-factor $g = 2$, as expected for an
electron bound to a hydrogenic donor potential.

Figure~\ref{fig3}(g) shows the Zeeman shifts of peaks $d_{1}$ and $d_{2}$
in barrier gate voltage. As expected, the peaks shifted linearly in
opposite directions, but the magnitudes of the shifts are different.
The peak $d_{1}$ shifted an average of -24 $\mu$VT$^{-1}$ whereas
the peak $d_{2}$ shifted an average of 95 $\mu$VT$^{-1}$, yielding
individual $g$-factors far from 2. However, if the difference of the
shifts for $d_{1}$ and $d_{2}$ is considered, we obtain the correct
value $g = 2.4 \pm 0.5$. The asymmetry in the peak shifts can be
caused, for example, by a slow global drift in the electric
potentials in the device over time.

While the nanoFET devices in this study were used to investigate
phosphorus donors, we note that these double-gated structures
could be used to study discrete energy states of any donor or
acceptor atom in silicon. In the latter case, this could be
achieved using a pMOS nanoFET, employing p$^{+}$ source--drain
contacts and a negative gate voltage. By employing on-chip
waveguides or microwave cavities to create a magnetic field in
resonance with the spin split $D^0$ donor state, one could use
these structures for the study of electrically-detected magnetic
resonance of single donor spins, as recently demonstrated for
small ensembles\cite{LoAPL2007}.

In conclusion, we have shown that resonant tunneling of electrons
through discrete energy states of phosphorus donors in silicon can
be observed in planar nanoscale FET structures, where the
presence and depth of the donors is deliberately chosen, and basic
device parameters such as tunnel couplings and electron density
can be conveniently controlled. We have identified the $D^0$ and $D^-$
charge states of individual donors, measured their charging energy,
and observed the Zeeman shift of their spin states in
magnetic fields up to 7~T. Control devices, deliberately
fabricated with no donors in the channel, showed markedly
different transport behavior than the characteristic features
present in donor-implanted devices. These experiments provide important
understanding of the electric and magnetic properties of
individual phosphorus donors needed for the realization of Si:P
nanoelectronics. \textcolor{black}{Moreover, these structures are compatible with single-ion implant detectors\cite{JamiesonAPL05,WeisJVSTB08} thus enabling devices with precisely one implanted donor.}
The demonstrated ability to resolve spin states of donors in these devices could also enable the measurement of electrically-detected magnetic resonance for a single-donor electron spin.

The authors thank D.\ Barber and R.\ P.\ Starrett for technical
support in the National Magnet Laboratory, UNSW, E.\ Gauja for
assistance in the UNSW Semiconductor Nanofabrication Facility, and A. Cimmino and R. Szymanski for technical support at the University of Melbourne. We also acknowledge J. McCallum,  L.\
C.\ L.\ Hollenberg, C.\ C.\ Escott and M.\ Eriksson for useful discussions.
M.\ M. and J.-M.\ P.\ acknowledge Academy of Finland
and Emil Aaltonen Foundation for financial support. This work is
supported by the Australian Research Council Centre of Excellence scheme, the Australian
Government, the U.S. National Security Agency (NSA), and the U.S.
Army Research Office (ARO) (under Contract No. W911NF-08-1-0527).

%\newpage


\begin{thebibliography}{37}
\expandafter\ifx\csname natexlab\endcsname\relax\def\natexlab#1{#1}\fi
\expandafter\ifx\csname bibnamefont\endcsname\relax
  \def\bibnamefont#1{#1}\fi
\expandafter\ifx\csname bibfnamefont\endcsname\relax
  \def\bibfnamefont#1{#1}\fi
\expandafter\ifx\csname citenamefont\endcsname\relax
  \def\citenamefont#1{#1}\fi
\expandafter\ifx\csname url\endcsname\relax
  \def\url#1{\texttt{#1}}\fi
\expandafter\ifx\csname urlprefix\endcsname\relax\def\urlprefix{URL }\fi
\providecommand{\bibinfo}[2]{#2}
\providecommand{\eprint}[2][]{\url{#2}}

\bibitem[{\citenamefont{Kane}({1998})}]{KaneNature98}
\bibinfo{author}{\bibfnamefont{B.}~\bibnamefont{Kane}},
  \bibinfo{journal}{{Nature}} \textbf{\bibinfo{volume}{{393}}},
  \bibinfo{pages}{133} (\bibinfo{year}{{1998}}).

\bibitem[{\citenamefont{Hollenberg et~al.}(2006)\citenamefont{Hollenberg,
  Greentree, Fowler, and Wellard}}]{hollenberg06PRB}
\bibinfo{author}{\bibfnamefont{L.~C.~L.} \bibnamefont{Hollenberg}},
  \bibinfo{author}{\bibfnamefont{A.~D.} \bibnamefont{Greentree}},
  \bibinfo{author}{\bibfnamefont{A.~G.} \bibnamefont{Fowler}},
  \bibnamefont{and} \bibinfo{author}{\bibfnamefont{C.~J.}
  \bibnamefont{Wellard}}, \bibinfo{journal}{Phys. Rev. B.}
  \textbf{\bibinfo{volume}{74}}, \bibinfo{eid}{045311} (\bibinfo{year}{2006}).

\bibitem[{\citenamefont{Lee et~al.}(2006)\citenamefont{Lee, Shin, Choi, Lee,
  Choi, Park, Yang, Lee, and Zyung}}]{LeeAPL06}
\bibinfo{author}{\bibfnamefont{S.~D.} \bibnamefont{Lee}},
  \bibinfo{author}{\bibfnamefont{S.~J.} \bibnamefont{Shin}},
  \bibinfo{author}{\bibfnamefont{S.~J.} \bibnamefont{Choi}},
  \bibinfo{author}{\bibfnamefont{J.~J.} \bibnamefont{Lee}},
  \bibinfo{author}{\bibfnamefont{J.~B.} \bibnamefont{Choi}},
  \bibinfo{author}{\bibfnamefont{S.}~\bibnamefont{Park}},
  \bibinfo{author}{\bibfnamefont{S.-R.~E.} \bibnamefont{Yang}},
  \bibinfo{author}{\bibfnamefont{S.~J.} \bibnamefont{Lee}}, \bibnamefont{and}
  \bibinfo{author}{\bibfnamefont{T.~H.} \bibnamefont{Zyung}},
  \bibinfo{journal}{Appl. Phys. Lett.} \textbf{\bibinfo{volume}{89}},
  \bibinfo{eid}{023111} (\bibinfo{year}{2006}).

\bibitem[{\citenamefont{Fujiwara et~al.}(2006)\citenamefont{Fujiwara, Inokawa,
  Yamazaki, Namatsu, Takahashi, Zimmerman, and Martin}}]{FujiwaraAPL06}
\bibinfo{author}{\bibfnamefont{A.}~\bibnamefont{Fujiwara}},
  \bibinfo{author}{\bibfnamefont{H.}~\bibnamefont{Inokawa}},
  \bibinfo{author}{\bibfnamefont{K.}~\bibnamefont{Yamazaki}},
  \bibinfo{author}{\bibfnamefont{H.}~\bibnamefont{Namatsu}},
  \bibinfo{author}{\bibfnamefont{Y.}~\bibnamefont{Takahashi}},
  \bibinfo{author}{\bibfnamefont{N.~M.} \bibnamefont{Zimmerman}},
  \bibnamefont{and} \bibinfo{author}{\bibfnamefont{S.~B.}
  \bibnamefont{Martin}}, \bibinfo{journal}{Appl. Phys. Lett.}
  \textbf{\bibinfo{volume}{88}}, \bibinfo{eid}{053121} (\bibinfo{year}{2006}).

\bibitem[{\citenamefont{Angus et~al.}(2007)\citenamefont{Angus, Ferguson,
  Dzurak, and Clark}}]{AngusNano07}
\bibinfo{author}{\bibfnamefont{S.~J.} \bibnamefont{Angus}},
  \bibinfo{author}{\bibfnamefont{A.~J.} \bibnamefont{Ferguson}},
  \bibinfo{author}{\bibfnamefont{A.~S.} \bibnamefont{Dzurak}},
  \bibnamefont{and} \bibinfo{author}{\bibfnamefont{R.~G.} \bibnamefont{Clark}},
  \bibinfo{journal}{Nano Letters} \textbf{\bibinfo{volume}{7}},
  \bibinfo{pages}{2051} (\bibinfo{year}{2007}).

\bibitem[{\citenamefont{Liu et~al.}(2008)\citenamefont{Liu, Fujisawa, Inokawa,
  Ono, Fujiwara, and Hirayama}}]{LiuAPL08}
\bibinfo{author}{\bibfnamefont{H.}~\bibnamefont{Liu}},
  \bibinfo{author}{\bibfnamefont{T.}~\bibnamefont{Fujisawa}},
  \bibinfo{author}{\bibfnamefont{H.}~\bibnamefont{Inokawa}},
  \bibinfo{author}{\bibfnamefont{Y.}~\bibnamefont{Ono}},
  \bibinfo{author}{\bibfnamefont{A.}~\bibnamefont{Fujiwara}}, \bibnamefont{and}
  \bibinfo{author}{\bibfnamefont{Y.}~\bibnamefont{Hirayama}},
  \bibinfo{journal}{Appl. Phys. Lett.} \textbf{\bibinfo{volume}{92}},
  \bibinfo{eid}{222104} (\bibinfo{year}{2008}).

\bibitem[{\citenamefont{Jamieson et~al.}(2005)\citenamefont{Jamieson, Yang,
  Hopf, Hearne, Pakes, Prawer, Mitic, Gauja, Andresen, Hudson
  et~al.}}]{JamiesonAPL05}
\bibinfo{author}{\bibfnamefont{D.~N.} \bibnamefont{Jamieson}},
  \bibinfo{author}{\bibfnamefont{C.}~\bibnamefont{Yang}},
  \bibinfo{author}{\bibfnamefont{T.}~\bibnamefont{Hopf}},
  \bibinfo{author}{\bibfnamefont{S.~M.} \bibnamefont{Hearne}},
  \bibinfo{author}{\bibfnamefont{C.~I.} \bibnamefont{Pakes}},
  \bibinfo{author}{\bibfnamefont{S.}~\bibnamefont{Prawer}},
  \bibinfo{author}{\bibfnamefont{M.}~\bibnamefont{Mitic}},
  \bibinfo{author}{\bibfnamefont{E.}~\bibnamefont{Gauja}},
  \bibinfo{author}{\bibfnamefont{S.~E.} \bibnamefont{Andresen}},
  \bibinfo{author}{\bibfnamefont{F.~E.} \bibnamefont{Hudson}},
  \bibnamefont{et~al.}, \bibinfo{journal}{Appl. Phys. Lett.}
  \textbf{\bibinfo{volume}{86}}, \bibinfo{eid}{202101} (\bibinfo{year}{2005}).

\bibitem[{\citenamefont{\textrm{A. Morello, C. C. Escott, H. Huebl, L. H.
  Willems van Beveren, L. C. L. Hollenberg, D. N. Jamieson, A. S. Dzurak, and
  R. G. Clark, arXiv:0904.1271} (unpublished)}()}]{MorelloPRB09}
\bibinfo{author}{\bibnamefont{\textrm{A. Morello, C. C. Escott, H. Huebl, L. H.
  Willems van Beveren, L. C. L. Hollenberg, D. N. Jamieson, A. S. Dzurak, and
  R. G. Clark, arXiv:0904.1271} (unpublished)}}.

\bibitem[{\citenamefont{Tyryshkin et~al.}(2003)\citenamefont{Tyryshkin, Lyon,
  Astashkin, and Raitsimring}}]{TyryshkinPRB03}
\bibinfo{author}{\bibfnamefont{A.~M.} \bibnamefont{Tyryshkin}},
  \bibinfo{author}{\bibfnamefont{S.~A.} \bibnamefont{Lyon}},
  \bibinfo{author}{\bibfnamefont{A.~V.} \bibnamefont{Astashkin}},
  \bibnamefont{and} \bibinfo{author}{\bibfnamefont{A.~M.}
  \bibnamefont{Raitsimring}}, \bibinfo{journal}{Phys. Rev. B}
  \textbf{\bibinfo{volume}{68}}, \bibinfo{pages}{193207}
  (\bibinfo{year}{2003}).

\bibitem[{\citenamefont{Schenkel et~al.}(2006)\citenamefont{Schenkel, Liddle,
  Persaud, Tyryshkin, Lyon, de~Sousa, Whaley, Bokor, Shangkuan, and
  Chakarov}}]{schenkel06APL}
\bibinfo{author}{\bibfnamefont{T.}~\bibnamefont{Schenkel}},
  \bibinfo{author}{\bibfnamefont{J.~A.} \bibnamefont{Liddle}},
  \bibinfo{author}{\bibfnamefont{A.}~\bibnamefont{Persaud}},
  \bibinfo{author}{\bibfnamefont{A.~M.} \bibnamefont{Tyryshkin}},
  \bibinfo{author}{\bibfnamefont{S.~A.} \bibnamefont{Lyon}},
  \bibinfo{author}{\bibfnamefont{R.}~\bibnamefont{de~Sousa}},
  \bibinfo{author}{\bibfnamefont{K.~B.} \bibnamefont{Whaley}},
  \bibinfo{author}{\bibfnamefont{J.}~\bibnamefont{Bokor}},
  \bibinfo{author}{\bibfnamefont{J.}~\bibnamefont{Shangkuan}},
  \bibnamefont{and} \bibinfo{author}{\bibfnamefont{I.}~\bibnamefont{Chakarov}},
  \bibinfo{journal}{Appl. Phys. Lett.} \textbf{\bibinfo{volume}{88}},
  \bibinfo{eid}{112101} (\bibinfo{year}{2006}).

\bibitem[{\citenamefont{de~Sousa}(2007)}]{desousa07PRB}
\bibinfo{author}{\bibfnamefont{R.}~\bibnamefont{de~Sousa}},
  \bibinfo{journal}{Phys. Rev. B.} \textbf{\bibinfo{volume}{76}},
  \bibinfo{eid}{245306} (\bibinfo{year}{2007}).

\bibitem[{\citenamefont{Sellier et~al.}(2006)\citenamefont{Sellier, Lansbergen,
  Caro, Rogge, Collaert, Ferain, Jurczak, and Biesemans}}]{SellierPRL06}
\bibinfo{author}{\bibfnamefont{H.}~\bibnamefont{Sellier}},
  \bibinfo{author}{\bibfnamefont{G.~P.} \bibnamefont{Lansbergen}},
  \bibinfo{author}{\bibfnamefont{J.}~\bibnamefont{Caro}},
  \bibinfo{author}{\bibfnamefont{S.}~\bibnamefont{Rogge}},
  \bibinfo{author}{\bibfnamefont{N.}~\bibnamefont{Collaert}},
  \bibinfo{author}{\bibfnamefont{I.}~\bibnamefont{Ferain}},
  \bibinfo{author}{\bibfnamefont{M.}~\bibnamefont{Jurczak}}, \bibnamefont{and}
  \bibinfo{author}{\bibfnamefont{S.}~\bibnamefont{Biesemans}},
  \bibinfo{journal}{Phys. Rev. Lett.} \textbf{\bibinfo{volume}{97}},
  \bibinfo{eid}{206805} (\bibinfo{year}{2006}).

\bibitem[{\citenamefont{Lansbergen et~al.}(2008)\citenamefont{Lansbergen,
  Rahman, Wellard, Woo, Caro, Collaert, Biesemans, Klimeck, Hollenberg, and
  Rogge}}]{LansbergenNatPhys08}
\bibinfo{author}{\bibfnamefont{G.~P.} \bibnamefont{Lansbergen}},
  \bibinfo{author}{\bibfnamefont{R.}~\bibnamefont{Rahman}},
  \bibinfo{author}{\bibfnamefont{C.~J.} \bibnamefont{Wellard}},
  \bibinfo{author}{\bibfnamefont{I.}~\bibnamefont{Woo}},
  \bibinfo{author}{\bibfnamefont{J.}~\bibnamefont{Caro}},
  \bibinfo{author}{\bibfnamefont{N.}~\bibnamefont{Collaert}},
  \bibinfo{author}{\bibfnamefont{S.}~\bibnamefont{Biesemans}},
  \bibinfo{author}{\bibfnamefont{G.}~\bibnamefont{Klimeck}},
  \bibinfo{author}{\bibfnamefont{L.~C.~L.} \bibnamefont{Hollenberg}},
  \bibnamefont{and} \bibinfo{author}{\bibfnamefont{S.}~\bibnamefont{Rogge}},
  \bibinfo{journal}{Nature Phys.} \textbf{\bibinfo{volume}{4}},
  \bibinfo{pages}{656} (\bibinfo{year}{2008}).

\bibitem[{\citenamefont{Calvet et~al.}(2008)\citenamefont{Calvet, Snyder, and
  Wernsdorfer}}]{CalvetPRB08}
\bibinfo{author}{\bibfnamefont{L.~E.} \bibnamefont{Calvet}},
  \bibinfo{author}{\bibfnamefont{J.~P.} \bibnamefont{Snyder}},
  \bibnamefont{and}
  \bibinfo{author}{\bibfnamefont{W.}~\bibnamefont{Wernsdorfer}},
  \bibinfo{journal}{Phys. Rev. B}  (\bibinfo{year}{2008}).

\bibitem[{\citenamefont{Calvet et~al.}(2007{\natexlab{a}})\citenamefont{Calvet,
  Wheeler, and Reed}}]{CalvetPRL07}
\bibinfo{author}{\bibfnamefont{L.~E.} \bibnamefont{Calvet}},
  \bibinfo{author}{\bibfnamefont{R.~G.} \bibnamefont{Wheeler}},
  \bibnamefont{and} \bibinfo{author}{\bibfnamefont{M.~A.} \bibnamefont{Reed}},
  \bibinfo{journal}{Phys. Rev. Lett.} \textbf{\bibinfo{volume}{98}},
  \bibinfo{pages}{096805} (\bibinfo{year}{2007}{\natexlab{a}}).

\bibitem[{\citenamefont{Calvet et~al.}(2007{\natexlab{b}})\citenamefont{Calvet,
  Wheeler, and Reed}}]{CalvetPRB07}
\bibinfo{author}{\bibfnamefont{L.~E.} \bibnamefont{Calvet}},
  \bibinfo{author}{\bibfnamefont{R.~G.} \bibnamefont{Wheeler}},
  \bibnamefont{and} \bibinfo{author}{\bibfnamefont{M.~A.} \bibnamefont{Reed}},
  \bibinfo{journal}{Phys. Rev. B} \textbf{\bibinfo{volume}{76}},
  \bibinfo{pages}{035319} (\bibinfo{year}{2007}{\natexlab{b}}).

\bibitem[{\citenamefont{\rm \textsc{T}opsil high-purity silicon (\textsc{HPS})
  wafers were~used [http://www.topsil.com].}()}]{Topsil}
\bibinfo{author}{\bibnamefont{\rm \textsc{T}opsil high-purity silicon
  (\textsc{HPS}) wafers were~used [http://www.topsil.com].}}

\bibitem[{\citenamefont{McCallum et~al.}(2008)\citenamefont{McCallum, Dunn, and
  Gauja}}]{McCallumMRSS08}
\bibinfo{author}{\bibfnamefont{J.}~\bibnamefont{McCallum}},
  \bibinfo{author}{\bibfnamefont{M.}~\bibnamefont{Dunn}}, \bibnamefont{and}
  \bibinfo{author}{\bibfnamefont{E.}~\bibnamefont{Gauja}},
  \bibinfo{journal}{Mater. Res. Soc. Symp. Proc.}
  \textbf{\bibinfo{volume}{1074}} (\bibinfo{year}{2008}).

\bibitem[{\citenamefont{Heij}(2001)}]{HeijPhD01}
\bibinfo{author}{\bibfnamefont{P.~C.} \bibnamefont{Heij}},
  \bibinfo{journal}{Ph.D. Thesis, Delft University of Technology} pp.
  \bibinfo{pages}{14--16} (\bibinfo{year}{2001}).

\bibitem[{\citenamefont{Lim et~al.}(2009)\citenamefont{Lim, Huebl, van Beveren,
  Rubanov, Spizzirri, Angus, Clark, and Dzurak}}]{LimAPL09}
\bibinfo{author}{\bibfnamefont{W.~H.} \bibnamefont{Lim}},
  \bibinfo{author}{\bibfnamefont{H.}~\bibnamefont{Huebl}},
  \bibinfo{author}{\bibfnamefont{L.~H.~W.} \bibnamefont{van Beveren}},
  \bibinfo{author}{\bibfnamefont{S.}~\bibnamefont{Rubanov}},
  \bibinfo{author}{\bibfnamefont{P.~G.} \bibnamefont{Spizzirri}},
  \bibinfo{author}{\bibfnamefont{S.~J.} \bibnamefont{Angus}},
  \bibinfo{author}{\bibfnamefont{R.~G.} \bibnamefont{Clark}}, \bibnamefont{and}
  \bibinfo{author}{\bibfnamefont{A.~S.} \bibnamefont{Dzurak}},
  \bibinfo{journal}{Appl. Phys. Lett.} \textbf{\bibinfo{volume}{94}},
  \bibinfo{eid}{173502} (\bibinfo{year}{2009}).

\bibitem[{Poi()}]{Poisson}
\emph{\bibinfo{title}{\rm \textsc{E}xplicitly, for $n$=3, we find
  \textsc{P}(0)=0.04, \textsc{P}(1)=0.14, \textsc{P}(2)=0.23,
  \textsc{P}(3)=0.24, \textsc{P}(4)=0.16, \textsc{P}(5)=0.10.}}

\bibitem[{FN1()}]{FN1}
\emph{\bibinfo{title}{\rm \textsc{T}his treatment neglects the mutual charging
  effects between the donor and other localized charges but should yield a good
  estimate here}}.

\bibitem[{\citenamefont{Ramdas and Rodriguez}(1981)}]{RamdasRPP81}
\bibinfo{author}{\bibfnamefont{A.~K.} \bibnamefont{Ramdas}} \bibnamefont{and}
  \bibinfo{author}{\bibfnamefont{S.}~\bibnamefont{Rodriguez}},
  \bibinfo{journal}{Reports on Progress in Physics}
  \textbf{\bibinfo{volume}{44}}, \bibinfo{pages}{1297} (\bibinfo{year}{1981}).

\bibitem[{\citenamefont{Taniguchi and Narita}(1976)}]{TaniguchiSSC76}
\bibinfo{author}{\bibfnamefont{M.}~\bibnamefont{Taniguchi}} \bibnamefont{and}
  \bibinfo{author}{\bibfnamefont{S.}~\bibnamefont{Narita}},
  \bibinfo{journal}{Solid State Communications} \textbf{\bibinfo{volume}{20}},
  \bibinfo{pages}{131 } (\bibinfo{year}{1976}).

\bibitem[{\citenamefont{\textrm{The software used to extract coupling
  capacitances is \textsc{Fastcap}, K. Nabors and J. White, IEEE Trans.
  Comput.-Aided Des. \textbf{10}, 1447.}}(1991)}]{Fastcap}
\bibinfo{author}{\bibnamefont{\textrm{The software used to extract coupling
  capacitances is \textsc{Fastcap}, K. Nabors and J. White, IEEE Trans.
  Comput.-Aided Des. \textbf{10}, 1447.}}} (\bibinfo{year}{1991}).

\bibitem[{SRI()}]{SRIM}
\emph{\bibinfo{title}{\rm \textsc{S}topping and range of ions in matter
  \textsc{(SRIM)} software was used to determine nominal phosphorous donor
  depth into silicon.}}

\bibitem[{FN4()}]{FN4}
\emph{\bibinfo{title}{\rm \textsc{D}ue to the large width of the closely
  adjacent peaks ($a_{2}$ and $c_{2}$) it was not possible to measure any shift
  with magnetic field, while peak $b_{1}$ showed an irregular non-monotonic
  shift.}}

\bibitem[{\citenamefont{Engel and Loss}(2001)}]{LossPRL01}
\bibinfo{author}{\bibfnamefont{H.-A.} \bibnamefont{Engel}} \bibnamefont{and}
  \bibinfo{author}{\bibfnamefont{D.}~\bibnamefont{Loss}},
  \bibinfo{journal}{Phys. Rev. Lett.} \textbf{\bibinfo{volume}{86}},
  \bibinfo{pages}{4648} (\bibinfo{year}{2001}).

\bibitem[{\citenamefont{Goldhaber-Gordon
  et~al.}(1998)\citenamefont{Goldhaber-Gordon, Shtrikman, Mahalu,
  Abusch-Magder, Meirav, and Kastner}}]{GordonNat98}
\bibinfo{author}{\bibfnamefont{D.}~\bibnamefont{Goldhaber-Gordon}},
  \bibinfo{author}{\bibfnamefont{H.}~\bibnamefont{Shtrikman}},
  \bibinfo{author}{\bibfnamefont{D.}~\bibnamefont{Mahalu}},
  \bibinfo{author}{\bibfnamefont{D.}~\bibnamefont{Abusch-Magder}},
  \bibinfo{author}{\bibfnamefont{U.}~\bibnamefont{Meirav}}, \bibnamefont{and}
  \bibinfo{author}{\bibfnamefont{M.~A.} \bibnamefont{Kastner}},
  \bibinfo{journal}{Nature} \textbf{\bibinfo{volume}{391}},
  \bibinfo{pages}{156} (\bibinfo{year}{1998}).

\bibitem[{\citenamefont{Hauptmann et~al.}(2008)\citenamefont{Hauptmann, Paaske,
  and Lindelof}}]{HauptmannNatPhys08}
\bibinfo{author}{\bibfnamefont{J.~R.} \bibnamefont{Hauptmann}},
  \bibinfo{author}{\bibfnamefont{J.}~\bibnamefont{Paaske}}, \bibnamefont{and}
  \bibinfo{author}{\bibfnamefont{P.~E.} \bibnamefont{Lindelof}},
  \bibinfo{journal}{Nat. Phys} \textbf{\bibinfo{volume}{4}},
  \bibinfo{pages}{373} (\bibinfo{year}{2008}).

\bibitem[{Rog()}]{Rogge09}
\emph{\bibinfo{title}{\rm \textsc{A}n extensive study of \textsc{K}ondo effects
  in silicon nano\textsc{MOSFET}s is in preparation by \textsc{G.P.
  L}ansbergen, \textsc{G.C. T}ettamanzi, \textsc{J. V}erduijn, \textsc{N.
  C}ollaert, \textsc{S. B}iesemans, \textsc{M. B}laauboer and \textsc{S.
  R}ogge.}}

\bibitem[{FN3()}]{FN3}
\emph{\bibinfo{title}{\rm \textsc{W}hen the resistances of the tunnel barriers
  are very high, any further widening of one of the barriers - due to the donor
  being off-center - would make the tunnel current immeasurably small.}}

\bibitem[{\citenamefont{\textrm{M. Pierre, M. Hofheinz, X. Jehl, M. Sanquer, G.
  Molas, M. Vinet, and S. Deleonibus, arXiv:0810.0672}}(2008)}]{Pierre08}
\bibinfo{author}{\bibnamefont{\textrm{M. Pierre, M. Hofheinz, X. Jehl, M.
  Sanquer, G. Molas, M. Vinet, and S. Deleonibus, arXiv:0810.0672}}}
  (\bibinfo{year}{2008}).

\bibitem[{\citenamefont{Xiao et~al.}(2004)\citenamefont{Xiao, Martin,
  Yablonovitch, and Jiang}}]{XiaoNat04}
\bibinfo{author}{\bibfnamefont{M.}~\bibnamefont{Xiao}},
  \bibinfo{author}{\bibfnamefont{I.}~\bibnamefont{Martin}},
  \bibinfo{author}{\bibfnamefont{E.}~\bibnamefont{Yablonovitch}},
  \bibnamefont{and} \bibinfo{author}{\bibfnamefont{H.}~\bibnamefont{Jiang}},
  \bibinfo{journal}{Nature} \textbf{\bibinfo{volume}{430}},
  \bibinfo{pages}{435} (\bibinfo{year}{2004}).

\bibitem[{FN2()}]{FN2}
\emph{\bibinfo{title}{\rm \textsc{A} detailed study of the mutual charging
  effects will be published elsewhere}}.

\bibitem[{\citenamefont{Lo et~al.}(2007)\citenamefont{Lo, Bokor, Schenkel,
  Tyryshkin, and Lyon}}]{LoAPL2007}
\bibinfo{author}{\bibfnamefont{C.~C.} \bibnamefont{Lo}},
  \bibinfo{author}{\bibfnamefont{J.}~\bibnamefont{Bokor}},
  \bibinfo{author}{\bibfnamefont{T.}~\bibnamefont{Schenkel}},
  \bibinfo{author}{\bibfnamefont{A.~M.} \bibnamefont{Tyryshkin}},
  \bibnamefont{and} \bibinfo{author}{\bibfnamefont{S.~A.} \bibnamefont{Lyon}},
  \bibinfo{journal}{Appl. Phys. Lett.} \textbf{\bibinfo{volume}{91}},
  \bibinfo{pages}{242106} (\bibinfo{year}{2007}).

\bibitem[{\citenamefont{Weis et~al.}(2008)\citenamefont{Weis, Schuh, Batra,
  Persaud, Rangelow, Bokor, Lo, Cabrini, Sideras-Haddad, Fuchs
  et~al.}}]{WeisJVSTB08}
\bibinfo{author}{\bibfnamefont{C.~D.} \bibnamefont{Weis}},
  \bibinfo{author}{\bibfnamefont{A.}~\bibnamefont{Schuh}},
  \bibinfo{author}{\bibfnamefont{A.}~\bibnamefont{Batra}},
  \bibinfo{author}{\bibfnamefont{A.}~\bibnamefont{Persaud}},
  \bibinfo{author}{\bibfnamefont{I.~W.} \bibnamefont{Rangelow}},
  \bibinfo{author}{\bibfnamefont{J.}~\bibnamefont{Bokor}},
  \bibinfo{author}{\bibfnamefont{C.~C.} \bibnamefont{Lo}},
  \bibinfo{author}{\bibfnamefont{S.}~\bibnamefont{Cabrini}},
  \bibinfo{author}{\bibfnamefont{E.}~\bibnamefont{Sideras-Haddad}},
  \bibinfo{author}{\bibfnamefont{G.~D.} \bibnamefont{Fuchs}},
  \bibnamefont{et~al.}, \bibinfo{journal}{J. Vac. Sci. Technol. B}
  \textbf{\bibinfo{volume}{26}}, \bibinfo{pages}{2596} (\bibinfo{year}{2008}).

\end{thebibliography}
\end{document}